\documentclass[english]{cccconf}
\usepackage[comma,numbers,square,sort&compress]{natbib}
\usepackage{epstopdf}
\usepackage{tabularx}
\usepackage{flushend}
\begin{document}

\title{A Data-Driven Vibration Analysis Framework for Micro-Motor Fault Diagnosis and Quality Control}

\author{Xuan Chen\aref{zju},
        Xinjun Zuo\aref{zju},
        Yancheng Bi\aref{zju},
        Shunli Yu\aref{sia},
        Yunqi Cao\aref{zju}}



\affiliation[zju]{College of Control Science and Engineering, Zhejiang University, Hangzhou, Zhejiang 310027, China
        \email{caoyunqi@zju.edu.cn}}
\affiliation[sia]{Ningbo Seago Intelligent Automation Co., Ltd., Ningbo, Zhejiang 315033, China
        \email{motoseago@seago.cn}}

\maketitle

\begin{abstract}
The reliability of the internal micro-motors is crucial for the performance and lifespan of electric toothbrushes. In this paper, a vibration-based fault detection method is proposed to identify micro-motor defects in electric toothbrushes. A dedicated signal acquisition device was designed and developed to capture the vibration signals of micro-motors using a high-precision accelerometer. To effectively characterize the micro-motor conditions, comprehensive features were extracted from the raw vibration data in both the time and frequency domains. A random forest (RF) algorithm was then employed to evaluate the importance of all extracted features. To better interpret the extracted features based on fault mechanisms, and to reduce dimensionality and computational overhead while avoiding overfitting, the top three features with the highest importance scores were selected to form the optimal feature subset. Finally, a support vector machine (SVM) model was utilized to classify the motor states based on the selected features. Experimental results demonstrate that the proposed method, combining RF-based feature selection and SVM classification, achieves outstanding diagnostic performance. Specifically, the model yields a balanced accuracy of 94.44\%, a defect recall of 88.89\%, a defect F1-score of 94.12\%, a Matthews correlation coefficient of 93.74\%, a geometric mean of 94.28\%, and an area under the receiver operating characteristic curve of 100.00\%. These robust metrics confirm that the proposed approach can accurately and efficiently detect micro-motor faults in electric toothbrushes, providing a practical and reliable solution for quality control and condition monitoring in manufacturing.
\end{abstract}


\footnotetext{This work was supported in part by the Open Research Project of the State Key Laboratory of Industrial Control Technology, China (Grant No. ICT2026B42), in part by Ningbo Seago Intelligent Automation Co., Ltd., and in part by Ningbo Seago Electric Co,.Ltd.}

\section{Introduction}
Electric toothbrushes have become an indispensable device in modern oral care, with their global market demand growing steadily due to increasing awareness of dental health \cite{yaacob2014powered}. The core driving component of an electric toothbrush is the internal micro-motor, which directly determines the device's cleaning efficiency, noise level, and operational lifespan. Consequently, the reliability of these micro-motors is paramount. In the manufacturing process, undetected defects such as rotor imbalance, bearing assembly errors, or winding inconsistencies can lead to excessive vibration and premature failure, significantly compromising product quality and brand reputation \cite{tavner2008review}. Therefore, developing an effective and non-destructive fault diagnosis method for micro-motors is crucial for quality control and condition monitoring in the consumer electronics industry \cite{lei2020applications}.

Traditionally, micro-motor defects in consumer electronics are identified through electrical current signature analysis or acoustic methods, which predominantly rely on manual auditory inspection \cite{jardine2006review, thomson2001current}. However, current-based methods often lack the sensitivity to identify subtle mechanical anomalies that do not significantly alter electrical parameters \cite{bellini2008advances}. Meanwhile, manual auditory inspection is highly subjective, labor-intensive, and prone to human error due to auditor fatigue and varying environmental background noise. In contrast, vibration-based condition monitoring offers a more robust alternative as it captures mechanical dynamics through direct contact \cite{tandon1999review}. With the aid of a dedicated acquisition device, the raw vibration signals obtained from micro-motors exhibit high stability and a favorable signal-to-noise ratio, effectively minimizing the interference commonly associated with airborne acoustic measurements. Despite the high quality of the acquired signals, extracting a comprehensive set of features from both time and frequency domains typically results in high-dimensional data. Directly utilizing all extracted features for classification can lead to increased computational overhead, the problem of high dimensionality, and a higher risk of overfitting, thereby reducing the model's generalization ability in real-world production environments \cite{gonzalez2021data}. Additionally, without a rigorous feature selection process, it remains difficult to interpret the extracted features in the context of specific physical fault mechanisms.

To address these challenges, this paper proposes a robust vibration-based fault diagnosis method specifically designed for electric toothbrush micro-motors. A dedicated signal acquisition device equipped with a high-precision accelerometer was developed to capture high-fidelity vibration data from the electric toothbrush housing, forming a specialized vibration acquisition system tailored for micro-motor testing. To effectively characterize the motor conditions, comprehensive features were extracted from the raw vibration data in both the time and frequency domains. Unlike conventional approaches that utilize all available features, this study employs a random forest (RF) \cite{breiman2001random} algorithm to evaluate the importance of all extracted features and selects the top three features with the highest importance scores to construct an optimal feature subset. This strategy not only reduces dimensionality and computational cost but also enhances the interpretability of the fault mechanisms by focusing on the most physically significant indicators. Finally, a support vector machine (SVM) \cite{boser1992training, cortes1995support} model is utilized to classify the micro-motor states based on this optimized subset, ensuring high diagnostic accuracy with minimal complexity. The proposed method was validated using a real-world dataset acquired from a manufacturing production line. Experimental results demonstrate robust performance in diagnosing micro-motor faults for electric toothbrushes, confirming the approach as a practical and reliable solution for quality control. Moreover, the methodology shows significant potential for extension to other similar consumer electronic products, underscoring its broader applicability.

\section{Methodology}
\subsection{Vibration Measurement System}
\begin{figure}[!t]
  \centering
  \includegraphics[width=\hsize]{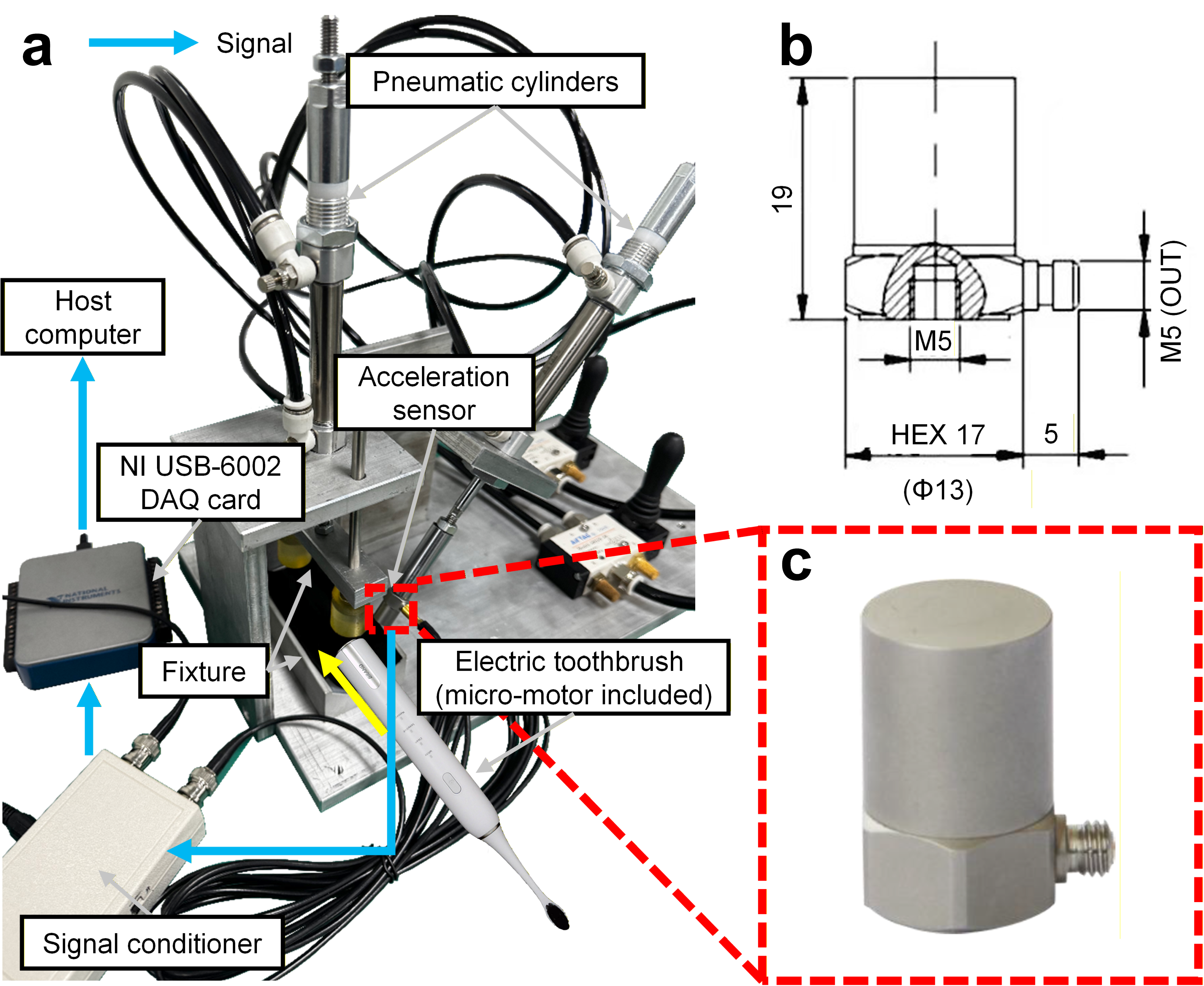}
  \caption{Design of the accelerometer-based vibration measurement system. (a) Experimental setup for vibration acquisition and analysis, showing a schematic of the measurement process with hardware connections and signal flow. (b) Dimensional drawing and (c) photograph of the employed acceleration sensor.}
  \label{fig:1}
\end{figure}

The vibration measurement system consists of pneumatic cylinders, an electric toothbrush, a fixture and the signal acquisition module, as shown in Fig.~\ref{fig:1}a. The signal acquisition module detects the vibration signal of the electric toothbrush mounted on the fixture using a high-precision acceleration sensor, as shown in Fig.~\ref{fig:1}b and c. The NI USB-6002 DAQ card is used to collect the output voltage signal from the acceleration sensor at a sampling frequency of $10~\mathrm{kHz}$. Leveraging the SVM classification model detailed in Section~\ref{sec:data processing and classification framework}, the host computer executes the micro-motor fault detection process and generates the final classification output.

\begin{figure}[!t]
  \centering
  \includegraphics[width=\hsize]{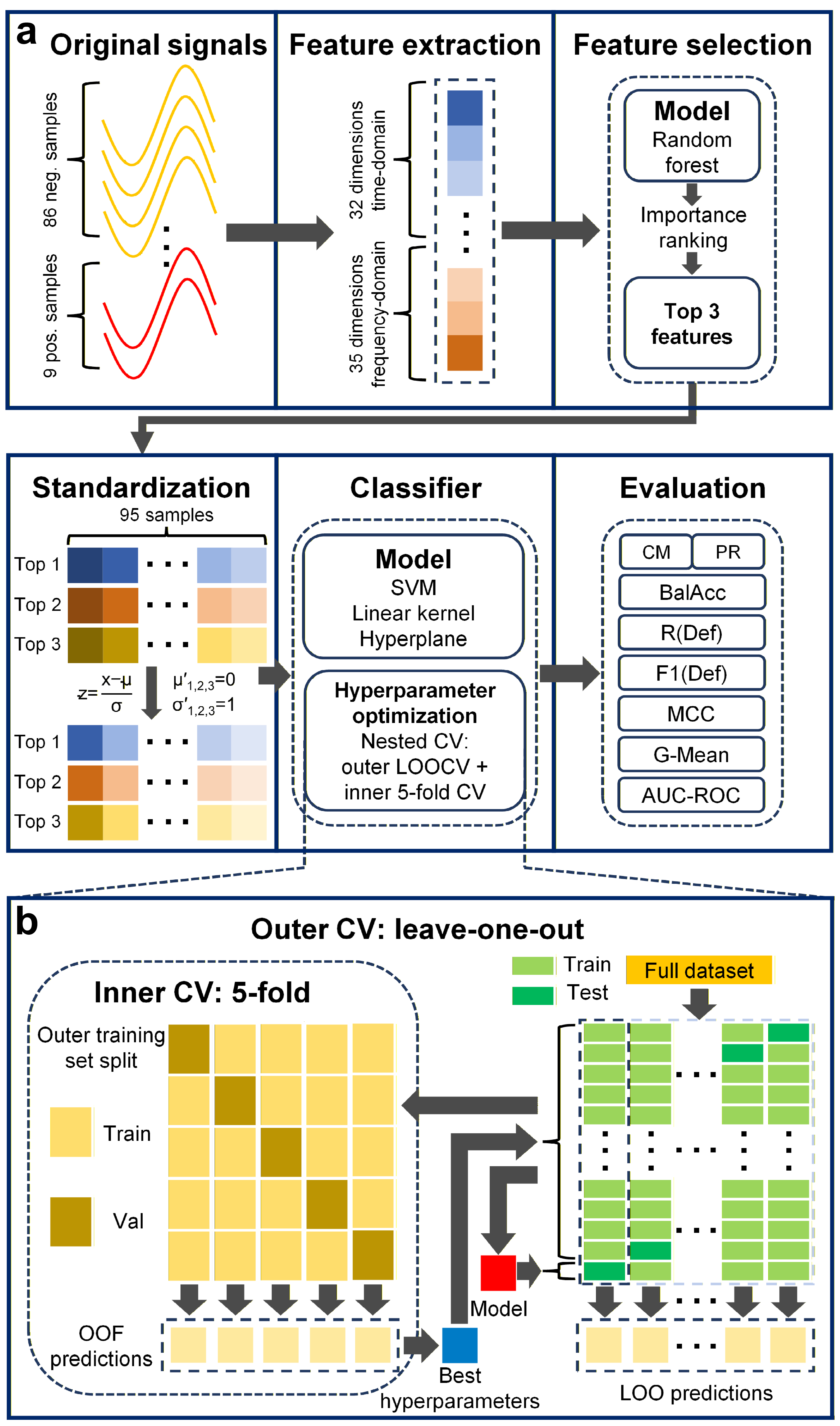}
  \caption{Flowchart of the proposed fault detection framework. (a) Overall workflow of data processing and classification, utilizing random forest for feature selection and support vector machine (SVM) for final classification. Evaluation metrics include confusion matrix (CM), precision-recall curve (PR), balanced accuracy (BalAcc), recall and F1-score for the defective class [R/F1(Def)], Matthews correlation coefficient (MCC), geometric mean (G-Mean), and area under the receiver operating characteristic curve (AUC-ROC). (b) Detailed schematic of the nested cross-validation (CV) strategy tailored for hyperparameter optimization and unbiased evaluation. The inner loop employs 5-fold CV for validation, while the outer loop utilizes leave-one-out CV (LOOCV) to generate out-of-fold (OOF) predictions.}
  \label{fig:2}
\end{figure}

\begin{figure*}[!t]
  \centering
  \includegraphics[width=\hsize]{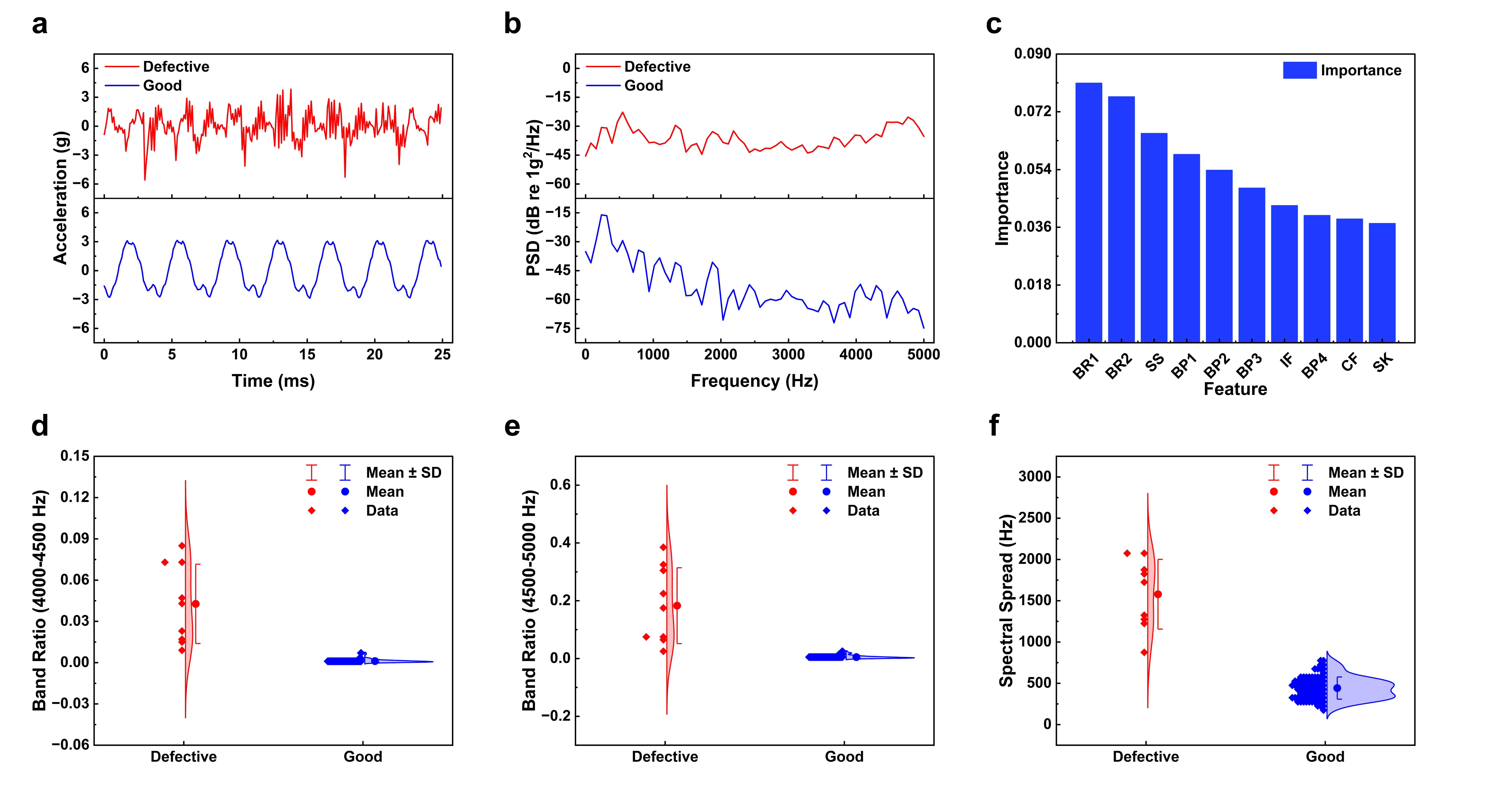}
  \caption{Vibration signal processing and feature engineering results. Comparison of (a) time-domain and (b) frequency-domain waveforms between defective and good acceleration signals, with power spectral density (PSD) estimated via Welch's method. (c) Top 10 feature importance ranking based on random forest Gini importance, including band ratio ($4000$--$4500~\mathrm{Hz}$) (BR1), band ratio ($4500$--$5000~\mathrm{Hz}$) (BR2), spectral spread (SS), band power ($4000$--$4500~\mathrm{Hz}$) (BP1), band power ($4500$--$5000~\mathrm{Hz}$) (BP2), band power ($2000$--$2500~\mathrm{Hz}$) (BP3), impulse factor (IF), band power ($2500$--$3000~\mathrm{Hz}$) (BP4), crest factor (CF), and spectral kurtosis (SK). Statistical comparison of selected spectral features between defective and good samples, specifically band ratios in the (d) $4000$--$4500~\mathrm{Hz}$ and (e) $4500$--$5000~\mathrm{Hz}$ frequency ranges, along with (f) spectral spread. Half-violin plots show data distribution with mean $\pm$ standard deviation (SD) indicated.}
  \label{fig:3}
\end{figure*}

\subsection{Data Processing and Classification Framework}
\label{sec:data processing and classification framework}
To automate the quality inspection and enable the diagnosis of micro-motor faults, this study addresses a highly imbalanced binary classification problem in which defective motor samples are defined as the positive class. Driven by the realities of factory production, where high manufacturing yields, the collected dataset exhibits an extreme class imbalance, containing $N_0$ good samples and $N_1$ defective samples ($N_0:N_1 = 86:9$). This severe skew underscores the critical need to minimize false negatives to prevent faulty products from reaching consumers. The dataset was collected from a continuous production run on the automated assembly line at Ningbo Seago Intelligent Automation Co., Ltd. To reflect realistic manufacturing yield rates while ensuring traceability, all 86 good samples and 9 defective samples were drawn from the same production batch. Defective units were pre-verified and labeled through standardized acoustic characteristic testing followed by a senior technician audit. This single-batch collection strategy was intentionally adopted to control for tooling wear and environmental process variations, thereby isolating fault-related vibrational signatures from batch-to-batch fluctuations while maintaining strict alignment with actual factory quality control workflows. Each sample is a univariate acceleration signal acquired at a sampling frequency of $10~\mathrm{kHz}$. To ensure feature extraction is performed on a common time scale and with consistent frequency resolution for accurate fault diagnosis, each raw signal is truncated or zero-padded to a fixed length of 50,000 points.

The workflow of the data processing and classification is shown in Fig.~\ref{fig:2}a. A comprehensive 67-dimensional feature set was extracted to characterize the vibrational signatures of the toothbrush motors, thereby capturing the underlying mechanical anomalies. This hybrid feature vector includes 32 time-domain statistical descriptors covering moment statistics, shape factors, Hjorth parameters, and percentile metrics, alongside 35 frequency-domain spectral features derived from Welch power spectral density (PSD) estimation that encompass spectral centroid, entropy, flatness, and energy distribution across ten contiguous $500~\mathrm{Hz}$ sub-bands within the Nyquist range. To mitigate the curse of dimensionality and enhance model interpretability, feature selection was performed using a RF classifier to rank all extracted features by Gini importance, with the top three most informative features retained for subsequent modeling.

Classification was implemented using a SVM with a linear kernel, deliberately selected over nonlinear alternatives or ensemble methods based on four key considerations. First, the selected high-frequency spectral features naturally induce a linearly separable representation of motor health states, rendering complex decision boundaries unnecessary. Second, given the constrained dataset size and severe class imbalance, a low-complexity linear model inherently mitigates overfitting risks compared to high-capacity nonlinear kernels or tree-based ensembles. Third, the linear SVM offers exceptional computational efficiency during inference, which is critical for real-time inline inspection at a 10 kHz sampling rate. Finally, the linear decision hyperplane yields transparent feature weights that directly map to physical fault mechanisms, satisfying the interpretability requirements of manufacturing quality control. This classifier was embedded within a pipeline that includes standardization to prevent data leakage during preprocessing. Model training and evaluation employed a leak-free nested cross-validation strategy, as shown in Fig.~\ref{fig:2}b, where the outer loop utilizes leave-one-out cross-validation (LOOCV) to provide an unbiased performance estimate, while the inner loop employs stratified 5-fold cross-validation for hyperparameter tuning of the regularization parameter and class weight optimization using balanced accuracy (BalAcc) as the scoring metric. This rigorous validation framework ensures that the test sample never participates in hyperparameter optimization, thereby yielding robust generalization performance metrics suitable for the imbalanced distribution of good and defective products. Final model evaluation incorporated comprehensive metrics including confusion matrix (CM), average precision (AP) from the precision-recall curve (PR), BalAcc, defect recall [R(Def)], defect F1-score [F1(Def)], Matthews correlation coefficient (MCC), geometric mean (G-mean), and area under the receiver operating characteristic curve (AUC-ROC) to provide a holistic assessment of classification performance under class imbalance.

\section{Results and Discussion}
\subsection{Signal Processing and Feature Engineering}
The vibration signal analysis revealed fundamental differences in dynamic behavior between defective and good toothbrush motors across both time and frequency domains, as shown in Fig.~\ref{fig:3}a and b. Time-domain examination demonstrated that defective products exhibited substantially higher vibration amplitudes with positive and negative peak accelerations of approximately $+4~\mathrm{g}$ and $-6~\mathrm{g}$, respectively, accompanied by irregular oscillation patterns indicative of mechanical instabilities such as bearing defects, rotor imbalances, or assembly misalignments. In contrast, good products displayed consistent, periodic waveforms with significantly lower amplitudes of approximately $\pm 3~\mathrm{g}$ and regular harmonic patterns reflecting stable motor operation. Frequency-domain analysis through Welch's PSD estimation further elucidated these distinctions, showing that defective motors exhibited elevated PSD levels across the entire $0$--$5000~\mathrm{Hz}$ spectrum ranging from $-45$ to $-20~\mathrm{dB}$ re $1~\mathrm{g^2/Hz}$, suggesting broadband vibration energy resulting from multiple defect mechanisms. Conversely, good motors demonstrated substantially lower PSD magnitudes of approximately $-75$ to $-15~\mathrm{dB}$ re $1~\mathrm{g^2/Hz}$ with distinct spectral peaks concentrated at specific operational frequencies, characteristic of normal harmonic behavior in well-manufactured micro-motors. This broadband versus narrowband energy distribution pattern provides a clear physical basis for distinguishing between good and defective products.

Feature importance analysis using RF classification identified that high-frequency spectral characteristics provide the most discriminative information for quality classification, with high-frequency band ratios in the $4000$--$5000~\mathrm{Hz}$ range and spectral shape descriptors predominating among top-ranked features, as shown in Fig.~\ref{fig:3}c. This finding aligns with vibration theory, as mechanical defects typically generate high-frequency components through impact phenomena, friction-induced vibrations, and structural resonance excitation \cite{randall2021vibration, tandon1999review, antoni2006spectral}. Statistical analysis revealed clear separation between classes, with defective products exhibiting band ratio values approximately $37$ times higher than good products in the $4000$--$5000~\mathrm{Hz}$ range, accompanied by minimal distributional overlap, as shown in Fig.~\ref{fig:3}d and e. The spectral spread feature further differentiated classes, with defective motors showing values approximately four times higher than good motors, specifically $1579 \pm 398~\mathrm{Hz}$ versus $442 \pm 134~\mathrm{Hz}$, as shown in Fig.~\ref{fig:3}f, reflecting broader frequency distribution of vibration energy in faulty products. The pronounced class separation in these high-frequency features, combined with their physical interpretability, validates the feature selection strategy and provides a robust foundation for supervised classification. These results suggest that effective quality discrimination can be achieved through targeted monitoring of the $4000$--$5000~\mathrm{Hz}$ frequency range, offering practical implications for real-time inspection systems while providing insights into critical manufacturing quality control points related to bearing condition, rotor balance, and assembly precision.

\subsection{Classification Performance of the SVM Model}
\begin{figure*}[!t]
  \centering
  \includegraphics[width=\hsize]{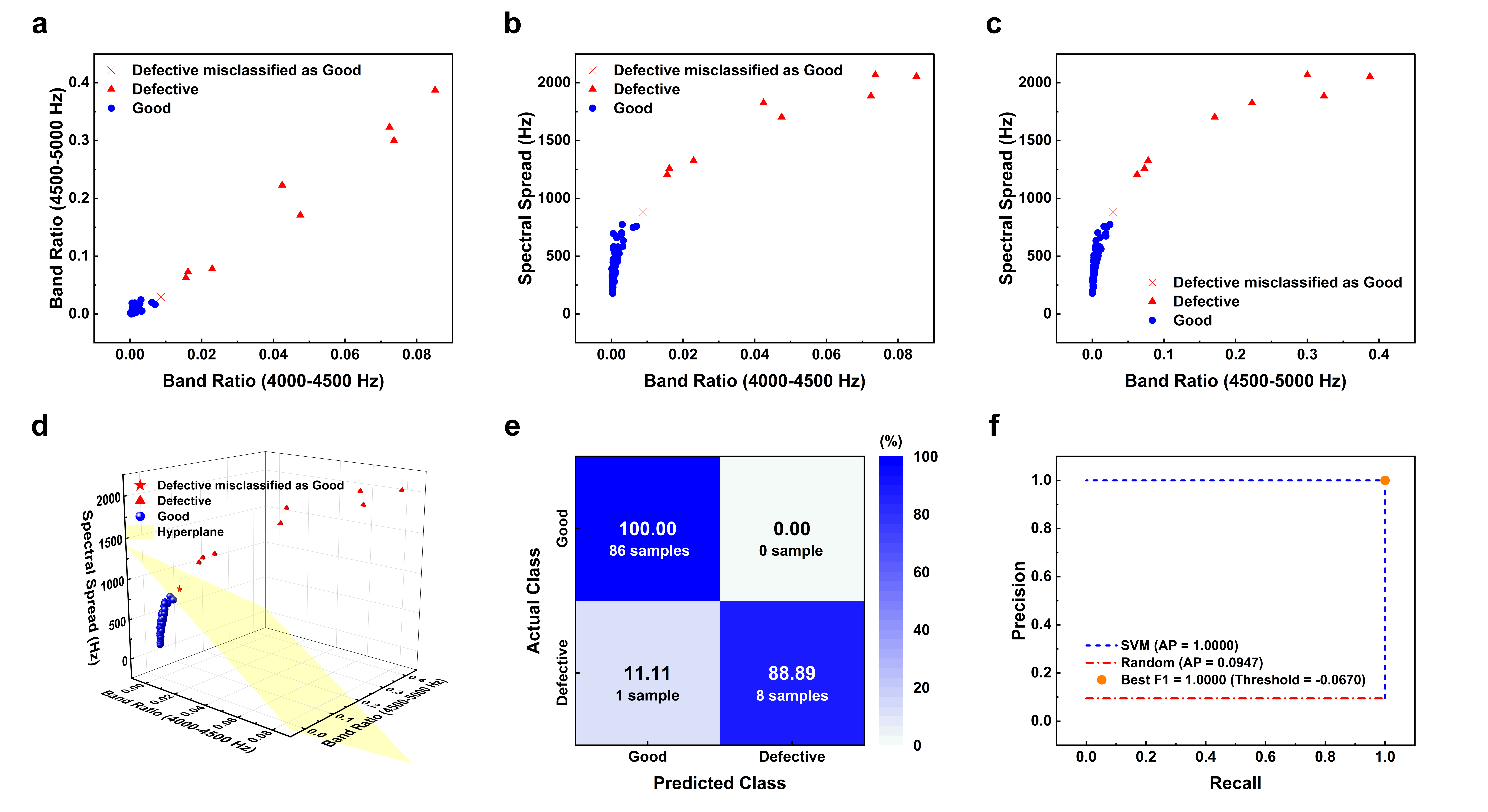}
  \caption{Classification results of the support vector machine (SVM) model based on the top three selected features and evaluated under the nested cross-validation strategy, illustrating (a)--(c) two-dimensional feature separability, (d) the three-dimensional feature distribution separated by the decision hyperplane, (e) the confusion matrix, and (f) the precision-recall (PR) curve with average precision (AP) metric.}
  \label{fig:4}
\end{figure*}

Two-dimensional feature space visualization revealed clear class separation along multiple feature combinations. The scatter plots in the band ratio planes, shown in Fig.~\ref{fig:4}a to c, demonstrated that defective products consistently exhibited higher band ratio values compared to good products, forming distinct clusters in the high-frequency energy distribution space. This separation pattern confirms that mechanical defects in toothbrush motors predominantly manifest as elevated high-frequency vibrational energy in the $4000$--$5000~\mathrm{Hz}$ range, thereby validating the feature selection strategy. The spectral spread feature further enhanced class discrimination, with defective motors displaying substantially broader spectral distributions with a mean of approximately $1579~\mathrm{Hz}$ compared to good motors with a mean of approximately $442~\mathrm{Hz}$, as evident in Fig.~\ref{fig:4}b and c. Three-dimensional visualization of the feature space, presented in Fig.~\ref{fig:4}d, illustrated the linear decision hyperplane learned by the SVM classifier. The optimal separating hyperplane effectively partitioned the feature space into two regions corresponding to good and defective classes, with a clear margin between the majority of samples. The geometric interpretation of the SVM decision boundary demonstrated that the selected features formed a linearly separable representation of motor quality, justifying the choice of linear kernel over more complex nonlinear alternatives. This linear separability suggests that the underlying physical mechanisms differentiating good and defective motors produce consistent, systematic variations in the vibrational signatures that can be captured by simple linear combinations of the selected features. The SVM classifier, trained on the top three selected features comprising band ratio $4000$--$4500~\mathrm{Hz}$, band ratio $4500$--$5000~\mathrm{Hz}$, and spectral spread, and evaluated using nested cross-validation, demonstrated excellent discriminative capability for toothbrush motor quality assessment. As summarized in Table~\ref{tab:1}, the overall classification performance achieved a BalAcc of 94.44\%, with perfect identification of good products at 100\% accuracy corresponding to 86 out of 86 samples, and high detection rate for defective products at 88.89\% accuracy corresponding to 8 out of 9 samples, as shown in Fig.~\ref{fig:4}e. The model also exhibited strong performance across complementary metrics, including a R(Def) of 88.89\%, F1(Def) of 94.12\%, MCC of 93.74\%, and G-Mean of 94.28\%, collectively indicating robust classification performance under the severe class imbalance condition with 9 defective versus 86 good samples.

\begin{table}[!htb]
  \centering
  \caption{Classification Performance of the SVM Model}
  \label{tab:1}
  \begin{tabularx}{0.85\linewidth}{@{\extracolsep{\fill}}l|l@{}}
    \hhline
    Metric & Value \\ \hline
    Balanced Accuracy (BalAcc) & $94.44\%$ \\ \hline
    Defect Recall [R(Def)] & $88.89\%$ \\ \hline
    Defect F1-score [F1(Def)] & $94.12\%$ \\ \hline
    Matthews Correlation Coefficient (MCC) & $93.74\%$ \\ \hline
    Geometric Mean (G-Mean) & $94.28\%$ \\ \hline
    Area Under ROC Curve (AUC-ROC) & $100.00\%$ \\
    \hhline
  \end{tabularx}
  \begin{minipage}{0.85\linewidth}
    \vspace{2pt}
    \footnotesize
    \textit{Note:} ROC denotes the Receiver Operating Characteristic curve.
  \end{minipage}
\end{table}

The confusion matrix analysis, depicted in Fig.~\ref{fig:4}e, revealed that the single misclassification occurred in the defective class, where one defective sample was incorrectly predicted as good, representing a false negative and a critical quality control concern as defective products escaping detection could reach end users. Examination of the misclassified sample's position in feature space, marked with a cross symbol in Fig.~\ref{fig:4}a to d, showed that it occupied an intermediate region between the good and defective clusters, exhibiting band ratio values lower than typical defective samples but spectral spread characteristics closer to the defective population. This transitional positioning indicates that the unit likely exhibits an early-stage or atypical defect mechanism, such as mild bearing wear or minor rotor eccentricity, which generates vibrational signatures that partially overlap with healthy operation. From a practical quality control perspective, this false negative highlights the importance of implementing appropriate decision thresholds that prioritize sensitivity, defined as defect detection rate, over specificity when the cost of defective products reaching customers is high. The precision-recall curve analysis, presented in Fig.~\ref{fig:4}f, yielded an AP of 1.0000, indicating that the model maintains perfect precision even when the decision threshold is adjusted toward higher recall to capture borderline defect cases with atypical severity, thereby confirming that the classification boundary can be adaptively tuned to recover marginal samples while preserving flawless positive prediction quality under severe class imbalance conditions, and substantially exceeding the random classifier baseline of 0.0947. Notably, the AUC-ROC achieved a perfect score of 100.00\% as shown in Table~\ref{tab:1}, further confirming the model's exceptional discriminative ability despite the limited defective sample size.

From a practical implementation perspective, the high classification accuracy, computational efficiency of linear SVM, and interpretable feature set make the proposed approach suitable for real-time quality inspection systems in toothbrush manufacturing, with the ability to detect defective motors with 88.89\% sensitivity while maintaining 100\% specificity for good products representing a favorable trade-off for production line deployment. However, the reported metrics should be interpreted with appropriate caution due to the limited number of defective samples. Small minority cohorts inherently introduce statistical variance in performance estimates and may occasionally yield optimistic generalization bounds when defect severity spans a broad spectrum. To enhance robustness and industrial readiness, future work will prioritize three directions. First, accumulating a larger, multi-batch dataset spanning different production shifts, tooling wear stages, and independent manufacturing lines to capture long-term process variations. Second, exploring signal-level data augmentation techniques such as time-warping, additive Gaussian noise, or physics-informed generative models to synthetically expand the minority class and stabilize decision boundaries. Third, validating the framework through cross-factory trials while implementing cost-sensitive thresholding or two-stage inspection protocols to minimize false negatives under real-world yield fluctuations. These steps will ensure the method remains reliable and scalable as manufacturing conditions evolve, ultimately bridging the gap between prototype validation and full-scale industrial deployment.

\section{Conclusion}
This paper presents a vibration-based fault diagnosis method for identifying micro-motor defects in electric toothbrushes, addressing the critical need for reliable quality control in manufacturing through a systematic approach combining RF-based feature selection and SVM classification. A dedicated signal acquisition system was developed using a high-precision accelerometer to capture vibration signatures, from which an optimal feature subset comprising band ratio in the $4000$--$4500~\mathrm{Hz}$ range, band ratio in the $4500$--$5000~\mathrm{Hz}$ range, and spectral spread was identified to effectively characterize high-frequency vibrational energy differences between defective and good motors. The proposed method achieved outstanding diagnostic performance with a BalAcc of 94.44\%, R(Def) of 88.89\%, F1(Def) of 94.12\%, MCC of 93.74\%, G-Mean of 94.28\%, and AUC-ROC of 100.00\%, demonstrating that a compact three-dimensional feature set can yield robust classification results under severe class imbalance conditions when evaluated through a rigorous nested cross-validation framework. Beyond the quantitative metrics, the physical interpretability of the selected features aligns with fault mechanisms involving bearing defects and rotor imbalances, while the computational efficiency of the linear SVM model enables potential deployment in real-time inline quality inspection systems. Future work will expand data collection across production batches and tooling wear stages, explore signal-level augmentation to enhance robustness under class imbalance, and integrate cost-sensitive thresholding with two-stage verification into an automated inspection system for reliable industrial deployment. Ultimately, this work provides an accurate, efficient, and interpretable solution for micro-motor quality assessment that contributes to improved product reliability and reduced manufacturing costs in the electric toothbrush industry.

\bibliographystyle{unsrt}
\bibliography{reference}

@article{yaacob2014powered,
  title={Powered versus manual toothbrushing for oral health},
  author={Yaacob, Munirah and Worthington, Helen V and Deacon, Scott A and Deery, Chris and Walmsley, A Damien and Robinson, Peter G and Glenny, Anne-Marie},
  journal={Cochrane Database of Systematic Reviews},
  number={6},
  year={2014},
  publisher={John Wiley \& Sons, Ltd}
}

@article{tavner2008review,
  title={Review of condition monitoring of rotating electrical machines},
  author={Tavner, Peter J},
  journal={IET electric power applications},
  volume={2},
  number={4},
  pages={215--247},
  year={2008},
  publisher={IET}
}

@article{lei2020applications,
  title={Applications of machine learning to machine fault diagnosis: A review and roadmap},
  author={Lei, Yaguo and Yang, Bin and Jiang, Xinwei and Jia, Feng and Li, Naipeng and Nandi, Asoke K},
  journal={Mechanical systems and signal processing},
  volume={138},
  pages={106587},
  year={2020},
  publisher={Elsevier}
}

@article{jardine2006review,
  title={A review on machinery diagnostics and prognostics implementing condition-based maintenance},
  author={Jardine, Andrew KS and Lin, Daming and Banjevic, Dragan},
  journal={Mechanical systems and signal processing},
  volume={20},
  number={7},
  pages={1483--1510},
  year={2006},
  publisher={Elsevier}
}

@article{thomson2001current,
  title={Current signature analysis to detect induction motor faults},
  author={Thomson, William T and Fenger, Mark},
  journal={IEEE Industry Applications Magazine},
  volume={7},
  number={4},
  pages={26--34},
  year={2001},
  publisher={Ieee}
}

@article{bellini2008advances,
  title={Advances in diagnostic techniques for induction machines},
  author={Bellini, Alberto and Filippetti, Fiorenzo and Tassoni, Carla and Capolino, G{\'e}rard-Andr{\'e}},
  journal={IEEE Transactions on industrial electronics},
  volume={55},
  number={12},
  pages={4109--4126},
  year={2008},
  publisher={IEEE}
}

@article{tandon1999review,
  title={A review of vibration and acoustic measurement methods for the detection of defects in rolling element bearings},
  author={Tandon, Naresh and Choudhury, Achintya},
  journal={Tribology international},
  volume={32},
  number={8},
  pages={469--480},
  year={1999},
  publisher={Elsevier}
}

@article{gonzalez2021data,
  title={Data-driven fault diagnosis for electric drives: A review},
  author={Gonzalez-Jimenez, David and Del-Olmo, Jon and Poza, Javier and Garramiola, Fernando and Madina, Patxi},
  journal={Sensors},
  volume={21},
  number={12},
  pages={4024},
  year={2021},
  publisher={MDPI}
}

@article{breiman2001random,
  title={Random forests},
  author={Breiman, Leo},
  journal={Machine learning},
  volume={45},
  number={1},
  pages={5--32},
  year={2001},
  publisher={Springer}
}

@inproceedings{boser1992training,
  title={A training algorithm for optimal margin classifiers},
  author={Boser, Bernhard E and Guyon, Isabelle M and Vapnik, Vladimir N},
  booktitle={Proceedings of the fifth annual workshop on Computational learning theory},
  pages={144--152},
  year={1992}
}

@article{cortes1995support,
  title={Support-vector networks},
  author={Cortes, Corinna and Vapnik, Vladimir},
  journal={Machine learning},
  volume={20},
  number={3},
  pages={273--297},
  year={1995},
  publisher={Springer}
}

@book{randall2021vibration,
  title={Vibration-based condition monitoring: industrial, automotive and aerospace applications},
  author={Randall, Robert Bond},
  year={2021},
  publisher={John Wiley \& Sons}
}

@article{antoni2006spectral,
  title={The spectral kurtosis: application to the vibratory surveillance and diagnostics of rotating machines},
  author={Antoni, J{\'e}r{\^o}me and Randall, Robert Bond},
  journal={Mechanical systems and signal processing},
  volume={20},
  number={2},
  pages={308--331},
  year={2006},
  publisher={Elsevier}
}

\end{document}